\documentclass[
aps,
pre,
preprint,
preprintnumbers,
tightenlines,
superscriptaddress,
amsmath,
amssymb,
noshowkeys,
nofootinbib,
floatfix,
]{revtex4-2}
\usepackage{graphicx}          
\usepackage{subfigure}         
\usepackage{dcolumn}           
\usepackage{capt-of}           
\usepackage{makecell}          
\usepackage{bbm}               
\usepackage{bm}                
\usepackage{time}              
\usepackage[mathlines]{lineno} 
\usepackage{stackrel}          
\usepackage{enumitem}          
\usepackage{physics}           
\usepackage[mathscr]{eucal}    
\usepackage{hyperref}          
\hypersetup{
    pdfnewwindow=true,         
    colorlinks=true,           
    linkcolor=red,             
    citecolor=blue,            
    filecolor=green,           
    urlcolor=cyan              
}
\usepackage{amsthm}
\theoremstyle{plain}                           

\theoremstyle{definition}

\theoremstyle{remark}

\newcommand{\tworule}{\vspace{\baselineskip}\noindent\hrule\vskip\doublerulesep\hrule}

\newcommand{\smod}{\text{S-S}}                 
\newcommand{\vmod}{\text{V-V}}                 
\newcommand{\ABS}{\text{ABS}}                  
\newcommand{\rmi}{{\rm i}}                     
\newcommand{\rme}{{\rm e}}                     
\newcommand{\Partial}[4]
   {\Bigl ( \frac{\partial #1 }{\partial #2 } \Bigr )_{\! #3, #4 }}
\newcommand{\tint}{\!\int\!}                   
\newcommand{\calL}{\mathcal{L}}

\newcommand{\calT}{\mathcal{T}}
\newcommand{\calV}{\mathcal{V}}
\newcommand{\bbR}{\mathbbm{R}}

\newcommand{\barPsi}{\bar{\Psi}}
\newcommand{\barPhi}{\bar{\Phi}}

\newcommand{\Tproduct}[1]%
   {\ensuremath{\mathcal{T} \{ \, #1 \, \} } }
\DeclareSymbolFont{matha}{OML}{txmi}{m}{it} 
\DeclareMathSymbol{\varv}{\mathord}{matha}{118}
\newcommand{\be}{\begin{equation}}
\newcommand{\ee}{\end{equation}}
\newcommand{\bea}{\begin{eqnarray}}
\newcommand{\eea}{\end{eqnarray}}

\begin{document}
%
%
\preprint{LAUR-}
\title{Solitary waves in a Two Parameter Family of Generalized  Nonlinear Dirac Equations in $1+1$ Dimensions}
\author{Avinash Khare}
\email[Email: ]{avinashkhare45@gmail.com} 
\affiliation{
   Physics Department, 
   Savitribai Phule Pune University, 
   Pune 411007, India} 
\author{Fred Cooper}
\email[Email: ]{cooper@santafe.edu}
\affiliation{
   Santa Fe Institute,
   1399 Hyde Park Road,
   Santa Fe, NM 87501, USA}
\affiliation{
   Center for Nonlinear Studies and Theoretical Division, 
   Los Alamos National Laboratory, 
   Los Alamos, NM 87545, USA}
\author{John F. Dawson}
\email[Email: ]{john.dawson@unh.edu}
\affiliation{
   Department of Physics,
   University of New Hampshire,
   Durham, NH 03824, USA}
%
\author{Avadh Saxena} 
\email[Email: ]{avadh@lanl.gov} 
\affiliation{
   Center for Nonlinear Studies and Theoretical Division, 
   Los Alamos National Laboratory, 
   Los Alamos, NM 87545, USA}
\date{\today, \now \ PST}
\begin{abstract}
We obtain exact solutions of the nonlinear Dirac equation in 1+1 dimension of the form  $  \Psi(x,t) =\Phi(x) \rme^{-\rmi \omega t}$
where the nonlinear interactions are a combination of vector-vector (V-V) and scalar-scalar (S-S) interactions with the interaction Lagrangian given by $L_I= \frac{g^2}{(\kappa+1)}(\bar{\psi} \psi)^{\kappa+1} 
-\frac{g^2}{p(\kappa+1)}[\bar{\psi} \gamma_{\mu} \psi \bar{\psi} \gamma^{\mu} 
\psi]^{(\kappa+1)/2}$. This generalizes the model of ABS (N.V. Alexeeva, I.V. Barashenkov and A. Saxena, Annals Phys. 
{\bf 403}, 198, (2019)) by having the arbitrary  nonlinearity parameter 
$\kappa>0$ and by replacing the coefficient of the V-V interaction by the 
arbitrary positive parameter $p>1$ which alters the relative weights of the 
vector-vector and the scalar-scalar interactions. We show that the solitary 
wave solutions exist in the entire allowed $(\kappa,p)$ plane for $\omega/m > 1/p^{1/(\kappa+1)} $, 
for frequency $\omega$ and mass $m$. These solutions have the property that their 
energy divided by their charge is $\it {independent} $ of the coupling constant
$g$. As $\omega$ increases, there is a transition from the double humped to the
 single humped solitons. We discuss the regions of stability of these solutions
 as a function of $\omega,p,\kappa$ using the Vakhitov-Kolokolov criterion. 
Finally we discuss the non-relativistic reduction of the 2-parameter family
of  generalized ABS models to a modified nonlinear Schr\"odinger equation 
(NLSE) and discuss the stability of the solitary waves in the domain of 
validity of the modified NLSE.
\end{abstract}
\maketitle
\tworule
\section{\label{s:Intro}Introduction}

In the last several years nonlinear Dirac equation (NLDE) has found novel 
application in condensed matter physics. For example, after the discovery of 
grapheme, a truly two-dimensional ($2D$) solid state material, it has been 
realized that the long wave length limit physics, especially on a honeycomb 
lattice is described by the massless Dirac equation \cite{3,4}. It has also been shown 
that in the mean field limit, the Bose-Einstein condensates on a honeycomb 
optical lattice are described by a NLDE \cite{R11}. 
Another area where NLDE has found application is in silicene transition metal 
dichalcogenides \cite{R9}. 
The recent applications of the Dirac equation in optics are to the 
light propagation in honeycomb photo-refractive lattices such as photonic 
grapheme \cite{R10} and conical diffraction in such structures \cite{R11}. The 
spin-orbit coupled Bose-Einstein condensates is yet another area that utilizes 
the nonlinear Dirac-type equations \cite{R12}. 

It is worth recalling here that the nonlinear Dirac equation has a very long 
history especially in the context of high energy physics. The nonlinear Dirac 
equation with the scalar-scalar (S-S) interaction was introduced by Ivanenko 
\cite{R1} as early as 1938 while corresponding one with the vector-vector (V-V) 
interaction was introduced by Thirring in 1958 \cite{R3}. In the 1970s, Soler 
invoked the Ivanenko model to describe extended nucleons \cite{R4}. The 
one dimensional version of the Soler model known as the Gross-Neveu model was 
introduced in 1974 to explain quark confinement \cite{R5}. Subsequently, its explicit 
solitary wave solutions were soon obtained \cite{lee}. Mathematically, the 
Thirring model was shown to be completely integrable via the inverse scattering
 transform \cite{mikh,kaup}. Outside the realm of elementary particles, the 
Gross-Neveu model was utilized in the study of conducting polymers \cite{bc} while the 
massive Thirring model found applications  in the context of optical gratings 
\cite{gratings}. 

A generalization of the  nonlinear Dirac equation with the S-S interaction 
\cite{R4}  and V-V interaction \cite{R3} to having 
arbitrary nonlinearity parameter $\kappa > 0$  was studied in 
\cite{fred}, where it  
was shown that while in the S-S case, the bound state solitary wave solutions 
exist for all values of $\kappa > 0$, for the V-V case such solutions are only 
allowed if $\kappa < \kappa_c$. Recently, Alexeeva, Barashenkov and Saxena 
(ABS) \cite{abs} introduced a novel NLDE model at $\kappa = 1$ 
which includes both the S-S and the V-V interactions. In the present paper, we further 
generalize the ABS  model to a two (continuous) parameter family of models 
characterized by the nonlinearity parameter $\kappa > 0$ as well as a parameter
$p > 1$ characterizing the arbitrary admixture of the V-V interaction vis a vis 
the S-S interaction. The Lagrangian density for the generalized ABS model is of 
the form:
$L = i\bar{\psi} (\gamma^{\mu} \partial_{\mu} -m) \psi + L_I$, 
where
$L_I =({g^2}/{(\kappa+1)})(\bar{\psi} \psi)^{\kappa+1}-({g^2}/{p(\kappa+1)})
[\bar{\psi} \gamma_{\mu} \psi \bar{\psi} \gamma^{\mu} \psi]^{(\kappa+1)/2}$. 

Note that the ABS model has $\kappa = 1$ and $p = 2$ in which case there is an
additional supersymmetry. In this paper we analyze the generalized ABS model in
detail and obtain exact solitary wave solutions in the entire $\kappa$-$p$ plane 
(with $\kappa > 0, p >1$). The plan of the paper is the following. In Sec II we
 write the equations of motion and the Lagrangian for this generalized \ABS ~ model. 
 In Sec. III we obtain rest frame solitary wave 
solutions of the form $\psi(x,t) = e^{-i\omega t} \psi(x)$ which are functions 
of the rest-frame frequency $\omega$ as well as $p, \kappa, g$. In this paper
we are interested in looking for solutions for which $0 < \omega < m$. In Sec. 
IV we obtain expressions for the charge $Q$ and the energy $E$ of the solitary 
wave solutions and discuss the various properties of these solutions. We show that 
unlike the S-S and the V-V cases, in this model the
frequency $\omega$ is restricted to $1/p^{1/(\kappa+1)} < \omega /m < 1$. 

Further, we discuss the transition from the double humped to the single 
humped solutions as a function of $\omega$. In particular we show that $E/Q$ is
 independent of the coupling $g^2$ and that the solitary bound state solutions 
exist (i.e. $E/Q < m$) in the entire $(\kappa$-$p)$ plane. Besides, when 
$\kappa > 2$, $E/Q$ has a maximum as a function of $\omega$ for all $p$. This 
is related to the instability of the solitary wave solutions in a regime of 
$\omega > \omega_c(\kappa,p)$. Finally, we study the stability of these bound 
states using the celebrated Vakhitov-Kolokolov \cite{vk} criterion and show
 the stability of these solitary bound states in case $\kappa \le 2$. In Sec. V
 we discuss the non-relativistic reduction of the generalized ABS model to the 
generalized modified nonlinear Schr\"odinger equation (NLSE) and discuss the 
stability of the solitary waves assuming the validity of the modified NLSE.
Finally, Sec. VI summarizes the results obtained in this paper and points out 
some of the open problems.
%
%
\section{\label{s:model}Single field massive nonlinear Dirac model}
A generalization of the  nonlinear Dirac equation for the Soler 
(\smod) \cite{R4} and Thirring (\vmod) \cite{R3} models to having arbitrary 
nonlinearity parameter $\kappa > 0$  was studied in \cite{fred}. That paper 
concluded that while the bound state solitary wave solutions exist in the 
entire $\kappa$ plane, in the V-V case such bound states are allowed only if 
$\kappa$ is less than a certain critical value. As noted above, the authors of Ref.  
(\ABS) \cite{abs} generalized the models discussed in \cite{fred} to
 include both the \vmod\ and \smod\ interactions for $\kappa=1$. In this paper,
 we further generalize this model to a two (continuous) parameter family of 
generalized ABS models characterized by the nonlinearity parameter $\kappa > 0$ and
a parameter $p > 1$ which alters the relative weight of the V-V interaction and 
the S-S interaction.

The Lagrangian density for this model is of the form:
\begin{align}\label{e:Lagrangian}
   \calL 
   &= 
   \barPsi \, [\, \rmi \gamma^{\mu} \partial_{\mu} - m \,] \, \Psi + \calL_I \>,
   \notag \\
   \calL_I
   &=
   \frac{g_s^2}{\kappa+1} \, [\barPsi \, \Psi]^{\kappa+1}
   +
   \frac{g_v^2}{\kappa+1} \, 
   [\, (\barPsi \gamma_{\mu} \Psi) (\barPsi \gamma^{\mu} \Psi) \,]^{(\kappa+1)/2} \,, 
\end{align}
which includes both Soler (\smod) \cite{R4} and Thirring (\vmod) \cite{R3} 
generalized interaction terms. The \smod\ case is when $g_s^2 = g^2$ and 
$g_v^2 = 0$, the \vmod\ case is when $g_s^2 = 0$ and $g_v^2 = g^2$, and the 
combined ABS case, is when $g_s^2 = g^2$ and $g_v^2 = - g^2/p < 0$, where 
$p = - g_s^2/g_v^2 > 1$ is a parameter we introduce to generalize the \ABS  
~model  \cite{abs} for which $\kappa = 1$  and $p=2$. 

The equation of motion for this combined model is:
\begin{equation}\label{e:absq-eom}
   (\rmi \gamma^{\mu} \partial_{\mu} - m) \Psi
   +
   g_s^2 (\barPsi \Psi)^{\kappa} \, \Psi
   +
   g_v^2 
   [ (\barPsi \gamma_{\nu} \Psi) (\barPsi \gamma^{\nu} \Psi) ]^{(\kappa-1)/2}
   (\barPsi \gamma_{\mu} \Psi)
   \gamma^{\mu} \Psi
   = 
   0 \>,
\end{equation}
and satisfies a current conservation equation,
\begin{equation}\label{e:CurrentConservation}
   \partial_{\mu} j^{\mu} = 0
   \qc
   j^{\mu} = \barPsi \, \gamma^{\mu} \Psi
   \qc
   Q = \int \dd[n]{x} \Psi^{\dag}(x) \Psi(x) \>.
\end{equation}
The energy-momentum tensor density also satisfies a conservation equation,
\begin{equation}\label{e:Tdef}
   \partial_{\mu} \calT^{\mu\nu}(x) = 0 
   \qc
   \calT^{\mu\nu} 
   =
   \rmi \, \barPsi \gamma^{\mu} \partial^{\nu} \Psi
   - 
   g^{\mu\nu} \calL  \>,
\end{equation}
which means that components of the linear 4-momentum vector $P^{\nu} = (E,\vb*{P})$ are conserved.  The conserved energy is given by
\begin{align}
   E
   &=
   \int \dd[d]{x} \calT^{0,0}
   =
   \int \dd[d]{x} 
   \qty\big[\,
      \rmi \, \barPsi \gamma^{0} \partial^{0} \Psi - \calL \,
   ]
   =
   \int \dd[d]{x} 
   \qty\big[\, 
      \barPsi \, \qty(\, \vb*{\gamma} \vdot \grad/\rmi + m \,)\, \Psi - \calV \,. 
   ]
   \label{e:calE}  
\end{align}
The ABS model is invariant under Lorentz transformations. The equations of motion can be derived from an Action Principle where the action 
$S$ is given by
\begin{equation}
S=\int \calL  d^n x dt; ~~  \calL
   =
   \barPsi (\rmi \gamma^{\mu} \partial_{\mu} - m) \Psi 
   +
   \calL_I \>,
\end{equation}
where 
 \begin{equation}\label{e:LIabspk}
  \calL_I
   =
 \frac{g^2}{(\kappa+1)}(\bar{\psi} \psi)^{\kappa+1} 
-\frac{g^2}{p(\kappa+1)} 
[\bar{\psi} \gamma_{\mu} \psi \bar{\psi} \gamma^{\mu} \psi]^{(\kappa+1)/2} .
\end{equation}

%
%
\section{\label{ss:SolitonSolution}Soliton solution}

In $1+1$ dimensions with $\dd{x^{\mu}} = ( \dd{t},\dd{x})$, we use the representations given in Ref.~\cite{fred} and choose
\begin{equation}\label{e.i:gammaReps}
   \gamma^{0} = \sigma_3 = 
   \mqty( 1 & 0 \\ 0 & -1 )
   \qc
   \gamma^{1} = \rmi \sigma_1 = 
    \rmi \mqty( 0 & 1 \\ 1 & 0 ) \>,
\end{equation}
where $\sigma_i$ are the Pauli matrices.  The gamma matrices then obey the anti-commutation relation, $\{ \gamma^{\mu}, \, \gamma^{\nu} \} = 2 g^{\mu\nu}$ as required.  Note also that
\begin{equation}\label{e.i:gammaProps}
   \gamma_{0} = \gamma^{0}
   \qc
   \gamma_{1} = - \gamma^{1}
   \qc
   \gamma^{0} \gamma^{1} 
   = 
   \rmi \, \sigma_3 \sigma_1 = - \sigma_2 = \rmi \, \vb{j}
   \qc
   \vb{j} = \mqty( 0 & 1 \\ -1 & 0 ) \>,
\end{equation}
where we have defined a new matrix $\vb{j} = \rmi \,\sigma_2$.  
So we seek solutions in the solitary wave rest frame of the form: 
\begin{equation}\label{e:Psi-2D}
   \Psi(x,t)
   =
   \Phi(x) \, \rme^{-\rmi \omega t}
   =
   \mqty( u(x) \\ v(x) ) \, \rme^{-\rmi \omega t}
   =
   R(x) \, \mqty( \cos[\theta(x)] \\ \sin[\theta(x)] ) \, 
   \rme^{-\rmi \omega t} \>,
\end{equation}
where $(R(x),\theta(x)) \in \bbR$ and $\omega$ is constrained  by requiring the
charge $Q$ to be finite. Moving solutions can be obtained from the rest frame 
solutions using the usual Lorentz boost due to the Lorentz invariance of the 
NLDE. 
  
From the equation of motion \eqref{e:absq-eom}, the soliton spinor $\Phi(x)$ satisfies
\begin{equation}\label{e:Phi-equ}
   \qty(\, \vb{j} \, \partial_x + \sigma_3 \, m \, ) \, \Phi
   -  
   g_s^2  (\barPhi \Phi)^{\kappa} \, \sigma_3 \, \Phi
   -
   g_v^2 
   [ (\barPhi \gamma_{\nu} \Phi) (\barPhi \gamma^{\nu} \Phi) ]^{(\kappa-1)/2}
   (\barPhi \gamma_{\mu} \Phi) \,
   \sigma_3 \, \gamma^{\mu} \Phi
   = 
   \omega \, \Phi \>.
\end{equation}
We define the potentials:
\begin{subequations}\label{e:VkUk-defs}
\begin{alignat}{4}
   V_k &= ( \Psi^T \sigma_3 \Psi )^{\kappa} 
    &= (u^2 - v^2)^{\kappa} &= R^{2\kappa} \cos^{\kappa}(2 \theta) \>,
   \label{e:Vk-def} \\
   U_k &= (\Psi^{T} \Psi )^{\kappa}
    &= (u^2 + v^2)^{\kappa} &= R^{2\kappa}  \>.
   \label{e:Uk-def}
\end{alignat}
\end{subequations}
Substitution of \eqref{e:Psi-2D}
into \eqref{e:absq-eom} yields in component form,
\begin{subequations}\label{e:uv-eom-I}
\begin{align}
   u' + ( m - g_s^2 V_{\kappa} ) \, v + ( \omega + g_v^2 U_{\kappa} ) \, v
   &= 0 \>,
   \label{e:uv-eom-I-a} \\  
   v' + ( m - g_s^2 V_{\kappa} ) \, u - ( \omega + g_v^2 U_{\kappa} ) \, u
   &= 0 \>. 
   \label{e:uv-eom-I-b}
\end{align}
\end{subequations}
Prime here means differentiation with respect to $x$. Multiplying 
\eqref{e:uv-eom-I-a} by $v$ and \eqref{e:uv-eom-I-b} by $u$, and adding and 
subtracting the two equations results in
\begin{subequations}\label{e:uv-add-subtract}
\begin{align}
   u v' + v u' 
   + 
   ( m - g_s V_{\kappa} ) \, (u^2 + v^2)
   -
   ( \omega + g_v^2 U_{\kappa} ) \, (u^2 - v^2)
   &=
   0 \>,
   \label{e:uv-add-subtract-a}
   \\
   u v' - v u' 
   + 
   ( m - g_s V_{\kappa} ) \, (u^2 - v^2)
   -
   ( \omega + g_v^2 U_{\kappa} ) \, (u^2 + v^2)
   &=
   0 \>.
   \label{e:uv-add-subtract-b}
\end{align}
\end{subequations}
In radial form, \eqref{e:uv-add-subtract-b} becomes:
\begin{equation}\label{e:uv-subtract-radial}
   \theta' 
   +  
   m  \cos(2\theta) - \omega 
   -
   g_s R^{2\kappa} \cos^{\kappa+1}(2\theta) )\,
   -
   g_v^2 R^{2 \kappa} = 0 \>.
\end{equation}
Multiplying \eqref{e:uv-eom-I-a} by $v0$ and \eqref{e:uv-eom-I-b} by $u$, we have  
\begin{subequations}\label{e:uuvv-II}
\begin{align}
   u u' + v v' + 2 \, ( m\, - g_s^2 V_{\kappa} ) \, u v
   &= 0 \>,
   \label{e:uuvv-II-a} \\
   u u' - v v' + 2 \, ( \omega + g_v^2 U_{\kappa} ) \, u v
   &= 0 \>.
   \label{e:uuvv-II-b}
\end{align}
\end{subequations}
Now multiply \eqref{e:uuvv-II-a} by $2 \, ( \omega + g_v^2 U_{\kappa} )$ and use \eqref{e:uuvv-II-b} to get
\begin{equation}\label{e:get-1}
   2 \, ( \omega + g_v^2 U_{\kappa} ) \, ( u u' + v v' )
   -
   2 \, ( m\, - g_s^2 V_{\kappa} ) \, ( u u' - v v' )
   =
   0 \>,
\end{equation}
which we can write as $\partial_x H(u,v) = 0$, where
\begin{equation}\label{e:Hu0v0-def}
   H(u,v)
   =
   \omega \, ( u^2 + v^2 )
   -
   m \, ( u^2 - v^2 )
   +
   \frac{g_v^2}{\kappa+1} U_{\kappa+1}
   +
   \frac{g_s^2}{\kappa+1} V_{\kappa+1} \,. 
\end{equation}
But as $|x| \rightarrow \infty$, for bound states, we require that both 
$(u(x),v(x)) \rightarrow 0$, which means that we can set $H(u,v) = 0$.  So then
in radial coordinates, \eqref{e:Hu0v0-def} gives:
\begin{equation}\label{e:Hcons-becomes}
   (\kappa+1) \, [\, \omega - m \cos(2\theta) \,]
   +
   g_v^2 \, R^{2\kappa}
   +
   g_s^2 R^{2\kappa} \cos^{\kappa+1}(2\theta)
   =
   0 \>.
\end{equation}
This same equation can be obtained using the fact that $\partial_x T^{11} = 0$.
Combining this with \eqref{e:uv-subtract-radial} gives a first order differential equation for $\theta(x)$:
\begin{equation}\label{e:theta-de}
   \frac{ d\theta }{dx} = \kappa \, [\, m \cos(2\theta) - \omega \, ] \>,
\end{equation}
which has the solution,
\begin{equation}\label{e:theta-sol}
   \theta(x) = \arctan[\, \alpha \tanh(\kappa \beta x ) \,] \>, 
\end{equation}
or equivalently
\be
\tan \theta =\alpha \tanh(\kappa \beta x ),
\ee
where
\begin{equation}\label{e:alphabeta}
   \alpha = \sqrt{\frac{m-\omega}{m+\omega}}
   \qc
   \beta = \sqrt{m^2 - \omega^2}.
\end{equation}
We require that $0 \le \omega \le m$. 
Solving \eqref{e:Hcons-becomes} for $R^{2\kappa}(x)$ gives
\begin{align}
   R^{2 \kappa}
   &=
   \frac{ (\kappa+1) \, [\, m \cos(2\theta) - \omega \,] }
        { g_s^2 \cos^{\kappa+1}(2\theta) + g_v^2 }
   =
   \frac{ (\kappa+1) \, p }{ g^2 } \,
   \qty[\,
   \frac{ m \cos(2\theta) - \omega }
        { p \cos^{\kappa+1}(2\theta) - 1 } \,] \>,
   \label{e:Requ-II}
\end{align}
since for our case, $g_s^2 = g^2$ and $g_v^2 = - g^2/p$. 
Real solutions $u(x)$ and $v(x)$ are then given by \eqref{e:Psi-2D}.
As expected, in the limit $p \rightarrow \infty$,  $R^2$ reduces to that of the
S-S case \cite{fred}. On using Eq. \eqref{e:theta-de} and Eq. \eqref{e:theta-sol}
 and the identities
\be\label{24}
m+\omega\cosh(2\kappa \beta x) = \frac{(m+\omega)
[1-\alpha^2 \tanh^2(\kappa \beta x)]}{\sech^2(\kappa \beta x)}\,,
\ee
\be\label{25}
\omega+m\cosh(2\kappa \beta x) = \frac{(m+\omega)
[1+\alpha^2 \tanh^2(\kappa \beta x)]}{\sech^2(\kappa \beta x)}\,,
\ee
one obtains an alternative expression for $R^2$ 
\be\label{26}
R^2= (1+\alpha^2 y^2) \bigg[\frac{p(\kappa+1)(1-y^2) (m-\omega)}{g^2 
(p(1- \alpha^2 y^2)^{\kappa+1}-(1+\alpha^2 y^2)^{\kappa+1})} \bigg] ^{1/\kappa} \,,  
\ee
where $y= \tanh (\kappa \beta x)$.

%
%
\section{Properties of the Solitary Wave Solutions}

\subsection{Double Hump and Single Hump regions of $ R^2$ as a function of $p,\kappa$. }
For the case $p\rightarrow\infty$ (scalar-scalar case)  there is a transition from double humped solitons to single humped solitons once 
\begin{equation}
1 \ge \frac{\omega}{m} > \frac{\kappa}{\kappa+1}. 
\end{equation}
We now derive such an expression in the present model with arbitrary $p,\kappa$. 
Observe that $R^2$ can be written in a factorized form as:
 \be
 R^2 = C(\omega,\kappa,p)
(1+\alpha^2 y^2)\frac{(1-y^2) ^{2/\kappa}}
{[p(1-\alpha^2 y^2)^{(\kappa+1)} 
- (1+\alpha^2 y^2)^{(\kappa+1)}]^{1/\kappa}}\,,
\ee
where
\be \label{C} 
 C(\omega,\kappa,p)
   =
   \frac{2 p^{1/\kappa}(\kappa+1)^{1/\kappa}
          \beta^{(1-\kappa)/\kappa} \alpha^{1/\kappa}}
        {\kappa g^{2/\kappa}}   \,. 
\ee
Letting  $S=(1-y^2) = \sech(\beta  \kappa x)$ 
we find 
\bea\label{3.5}
&&\frac{dR^2(x)/C}{dx} = 2 y S^{2/\kappa}\beta \frac{[\alpha^2 S^{2} -(1/\kappa)
(1+\alpha^2 y^2)]}{[p(1-\alpha^2 y^2)^{(\kappa+1)} 
- (1+\alpha^2 y^2)^{(\kappa+1)}]^{1/\kappa}} \nonumber \\
&&+\frac{2(\kappa+1)\beta\alpha^2}{\kappa} S^{2(\kappa+1)/\kappa} y 
\frac{p(1-\alpha^2 y^2)^{\kappa}-(1+\alpha^2 y^2)^{\kappa}}
{[p(1-\alpha^2 y^2)^{(\kappa+1)} 
- (1+\alpha^2 y^2)^{(\kappa+1)}]^{(\kappa+1)/\kappa}}\,.
\eea

\begin{figure}[t]
   \centering
   \includegraphics[width=0.75\linewidth]{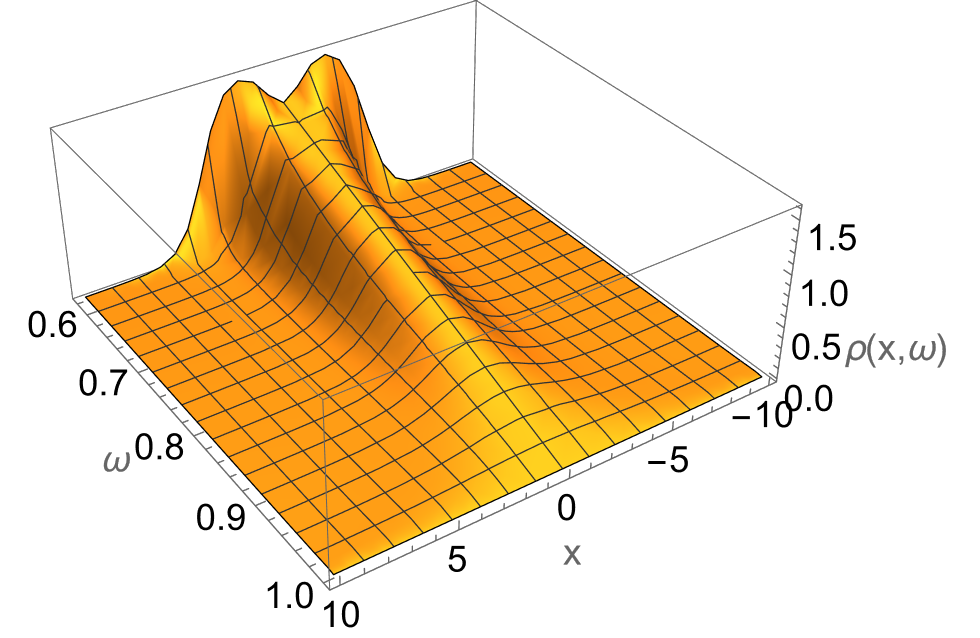}
   \caption{\label{f:Fig3}Plot of the charge density $\rho(x,\omega)$
   for $\kappa=1$, $g=1$, and $p=3$.}
\end{figure}

The minima of $R^2$ at $\sech(\beta \kappa  x) = 0$, are  at 
$x = \pm \infty$. On the other hand $\tanh(\beta \kappa x) = 0$ is a maximum or a 
minimum of $R^2(x)$ depending on whether
\be\label{3.6}
[2p+\frac{(p+1)}{\kappa}]\alpha^2 - \frac{(p-1)}{\kappa} < (> ) ~0.
\end{equation} 
 This implies that
$\tanh(\beta \kappa  x) = 0$, i.e. $x = 0$ is a maximum of $R^2(x)$ in case 
$\omega/m > \frac{(p\kappa+1)}{p(\kappa+1)}$ while $x = 0$ is a minimum of 
$R^2(x)$ in case $\omega/m \le \frac{(p\kappa+1)}{p(\kappa+1)}$. 
In case both $x = 0$ and $x = \pm \infty$ are minima of
$R^2$, it follows that there must be two symmetric maxima in between.
Summarizing, one has a double hump in case 
\be\label{3.7}
\frac{1}{p^{1/(\kappa+1)}} < \frac{\omega}{m} \le \frac{(p\kappa+1)}
{p(\kappa+1)}\,,
\ee
while it has a single hump in case 
\be\label{3.8}
\frac{(p\kappa+1)}{p(\kappa+1)} < \frac{\omega}{m} < 1\,.
\ee
The transition from two humps to one hump as we increase $\omega$ 
when $p=3, \kappa=1$ 
is shown in Fig. (\ref{f:Fig3}).

\subsection{\label{s:Charge}Evaluation of Charge $Q$ and Restriction on 
$\omega$}

The charge is given by 
\begin{equation}\label{e:Q-I}
   Q(\omega,\kappa,p)
   =
   \int_{-\infty}^{\infty} \!\!\!\! \dd{x} | \Psi(x,t) |^2
   =
   \int_{-\infty}^{\infty} \!\!\!\! \dd{x} | R(x) |^2 \,. 
\end{equation}
 Setting $y = \tanh(\kappa \beta x)$,
Eq.  \eqref{e:Q-I} can be written as:
\begin{subequations}\label{e:QI-defs}
\begin{align}
   Q(\omega,\kappa,p)
   &=
   C(\omega,\kappa,p) \, I(\omega,\kappa,p)
        \qc
C(\omega,\kappa,p) = \frac{2}{\beta \kappa}
[\frac{p(\kappa+1)(m-\omega)}{g^2}]^{1/\kappa}
   \label{e:Q-def} \\
   I(\omega,\kappa,p)
   &=
   \int_{0}^{1} \dd{y} 
   \frac{(\, 1 + \alpha^2 \, y^2 \,)}{(\, 1 - y^2 \,)^{1-1/\kappa} \,
   \qty[\, 
         p (\, 1 - \alpha^2 \, y^2 \,)^{\kappa+1} 
         - 
         (\, 1 + \alpha^2 \, y^2 \,)^{\kappa+1}\,
       ]^{1/\kappa} } \>.
   \label{e:I-def}
\end{align}
\end{subequations}
Here $ C(\omega,\kappa,p)$ is defined in Eq. (\ref{C}).
Analytic forms can be found separately for the \smod\ and \vmod\ models, and 
agree with the earlier results \cite{fred}.  But for the general \ABS\ model, 
the evaluation of the integral requires a numerical calculation.  For the 
integral \eqref{e:I-def} to converge, the denominator must not vanish at $y=1$.
This in turn implies that $\omega$ is limited to
the region defined by 
\be\label{3.3}
\frac{1}{p^{1/(\kappa+1)}} < \frac{\omega}{m} < 1\,.
\ee

%
\subsection{The Total Energy $E$ and $g$ Independence of $E/Q$}
For the energy, from \eqref{e:calE} 
\begin{equation}\label{e:E-equation}
   E(\omega,\kappa,p)
   =
   \int_{-\infty}^{\infty} \!\!\! \dd{x}
   \Bigl \{\,
      \Psi^{T} \, 
         [\, \vb{j} \, \partial_x + \sigma_3 \, m \,] \, 
      \Psi
      -
      \frac{g_s^2}{\kappa+1} \, V_{\kappa+1}
      -    
      \frac{g_v^2}{\kappa+1} \, U_{\kappa+1}
   \Bigr \} \>.
\end{equation}
We can write this as
\begin{equation}
E(\omega,\kappa,p) = H_1+H_2-H_3 \,, 
\end{equation} 
where
\begin{eqnarray}
H_1&&=  \int_{-\infty}^{\infty} dx R^2(x) \frac{d\theta}{dx} ,~~H_2  = m\int_{-\infty}^{\infty} R^2(x) \cos(2\theta)\, dx , 
 \nonumber \\
H_3 &&= \frac{g^2}{\kappa+1} \int_{-\infty}^{\infty} dx R^{2\kappa} ( cos^{\kappa+1} (2 \theta) -\frac{1}{p}) \,. 
\end{eqnarray} 
Using
\begin{equation} \label{iden}
\frac{d\theta}{dx} = \kappa [m\cos(2\theta) -\omega]\,, ~~
\kappa H_3 = H_1,
 \end{equation}
we can  write:
 \begin{equation}
E(\omega,\kappa,p) = H_1(1-\frac{1}{\kappa}) + H_2 \,, 
 \end{equation}
where from Eq. (\ref{iden})
\begin{equation}
H_1=\kappa H_2 -\kappa \omega Q \,. 
 \end{equation} 
Therefore we obtain
\be \label{Ekp}
E(\omega,\kappa,p) = \kappa  H_2 -(\kappa-1) \omega Q\,,
\ee
where
\begin{subequations}\label{e:WJ-defs}
\begin{align}
   H_2
   &=
   m \int_{-\infty}^{\infty} \!\!\! \dd{x} R^2(x) \cos(2 \theta)
   = \frac{2m}{\kappa \beta} \,[\frac{p (\kappa+1)(m-\omega)}{g^2}]^{1/\kappa}
   \, J(\omega,\kappa,p)
   \label{e:W-def} \>,
   \\
   J(\omega,\kappa,p)
   &=
   \int_{0}^{1} \dd{y} 
   \frac{(\, 1 - \alpha^2 \, y^2 \,)}{(\, 1 - y^2 \,)^{1-1/\kappa} \,
   \qty[\, 
         p (\, 1 - \alpha^2 \, y^2 \,)^{\kappa+1} 
         - 
         (\, 1 + \alpha^2 \, y^2 \,)^{\kappa+1}\,
       ]^{1/\kappa} } \>.
   \label{e:J-def}
\end{align}
\end{subequations}

Notice that $H_2 \propto (g^2)^{-1/\kappa}$ and since as shown above (see Eq. \eqref{C}) 
$Q$ $\propto (g^2)^{-1/\kappa}$, we find  remarkably that  $H_2/Q$ and hence
$E/Q$ is independent of $g^2$ and only depends on $\omega, \kappa$ and $p$.
Unfortunately, $Q$ and $H_2$ have to be computed numerically except when
$\kappa = 1$.

%
%
\subsection{$E$ and $Q$ when $\kappa = 1$} 
When $\kappa = 1$, it follows from Eq. (\ref{Ekp}) that $H_2 = E$. At 
$\kappa = 1$, it follows from Eqs. \eqref{e:QI-defs} that
\be\label{3.18}
Q = \frac{4p\alpha}{g^2} \int_{0}^{1} dy~  \frac{(1+ \alpha^2 y^2)}
{[p (1-\alpha^2 y^2)^2 - (1+\alpha^2 y^2)^2]}\,.
\ee
Using partial fractions it is straightforward to evaluate $Q$ and we obtain
\be\label{3.19}
Q = \frac{2p}{g^2\sqrt{p-1}} \tanh^{-1}\big(\frac{\beta}{\omega\sqrt{p-1}}\big)\,.
\ee
Similarly, when $\kappa = 1$, it follows from Eqs. (\ref{Ekp}) and 
(\ref{e:WJ-defs}) that $H_2$ and hence total energy $E$ is given by 
\be\label{3.20}
E = H_2 =  \frac{4pm\alpha}{g^2} \int_0^1 ~dy~  \frac{(1-\alpha^2 y^2)}
{[p (1-\alpha^2 y^2)^2 - (1+\alpha^2 y^2)^2]}\,.
\ee
Using partial fractions it is straightforward to evaluate $H_2$ and we obtain
\be\label{3.21}
E = H_2 = \frac{2\sqrt{p}m}{g^2\sqrt{p-1}} \tanh^{-1}
\big(\frac{\sqrt{p} \beta}{m\sqrt{p-1}}\big)\,.
\ee
Notice that while $E$ and $Q$ are proportional to $g^{-2}$, their ratio $E/Q$ 
is independent of $g$. 

\begin{figure}[t]
    \centering
    \subfigure[\ $\omega(g,p)$, $Q=1$]
    {\label{f:Fig1a}
    \includegraphics[width=0.45\linewidth]{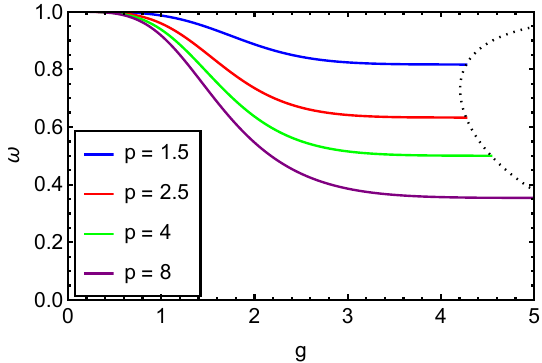}}
    \subfigure[\ $g(\omega,p)$, $Q=1$]
    {\label{f:Fig1b}
    \includegraphics[width=0.45\linewidth]{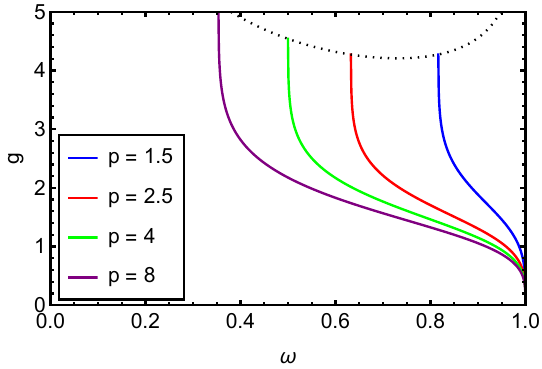}}
    \caption{\label{f:Fig1} Plots of $\omega(g,p)$ and 
    $g(\omega,p)$ for $\kappa=1$ when the charge $Q=1$.  The dotted
    lines indicate the soliton domain.}
\end{figure}

If we fix $Q$ to be a constant, say $Q=1$ we obtain an expression for $g^2$ as a function of $\omega$ and $p$:
\begin{equation}\label{e:gfunc}
g^ 2  = \frac{2p}{\sqrt{p-1}} \tanh^{-1}\big(\frac{\beta}{\omega\sqrt{p-1}}\big)\,.
\end{equation}
Similarly, inverting \eqref{e:gfunc}  gives an expression for $\omega$ in terms of $g$ and $p$:
\begin{equation}\label{e:CaseI-Q1-omega}
   \Bigl ( \frac{\omega}{m} \Bigr )^2
   =
   \frac{1}{1 + (p-1) \tanh^2[\, g^2\sqrt{p-1}/(2 p) \,]} \>.
\end{equation}
Plots of $\omega(g,p)$ and $g(\omega,p)$ for $\kappa=1$ when the charge $Q=1$ are shown in Fig.~\ref{f:Fig1}.  

The ratio $E/Q$  is however independent of $g$ and is given by 
\begin{equation} \label{EQ}
E/Q = \frac{m}{\sqrt{p}}\frac{  \tanh^{-1}
\big(\frac{\sqrt{p} \beta}{m\sqrt{p-1}}\big)}{ \tanh^{-1}\big(\frac{\beta}{\omega\sqrt{p-1}}\big) }  \,.
\end{equation} 

\begin{figure}[t]
    \centering
    \includegraphics[width=0.7 \linewidth]{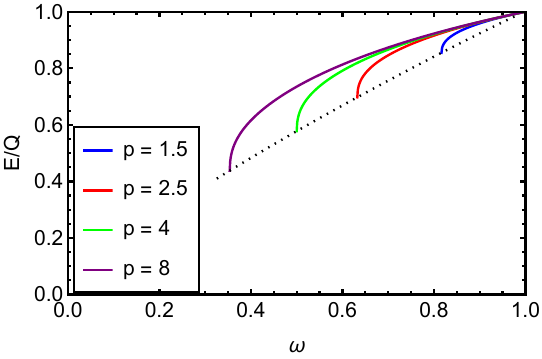}
    \caption{\label{f:Fig2} Plot of $E(\omega,p)/Q(\omega,p)$ when  $\kappa=1$. The dotted
    lines indicate the soliton domain.}
\end{figure}

From now onwards we set $m = 1$ so that $0 < \omega < 1$.
A plot of $E/Q$ as a function of  $\omega,p$ for $\kappa=1$  is shown in 
Fig.~\ref{f:Fig2}.  

%
%
%
%
%
%

%
%
%
%

%
%
\subsection{\label{ss:caseII}Case II: arbitrary \texorpdfstring{$\vb*{\kappa}$}{kappa2}}

\begin{figure}[t]
    \centering
    \subfigure[\ $g^2(\omega,\kappa,p)$, $p=2$]
    {\label{f:Fig4a}
    \includegraphics[width=0.45\linewidth]{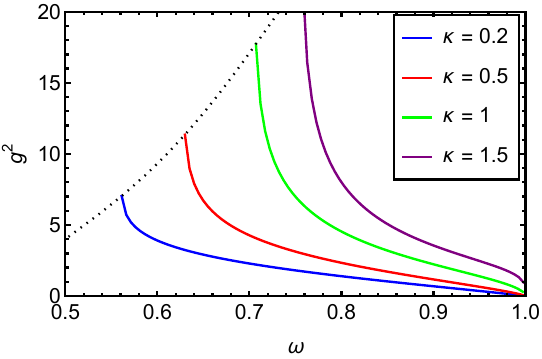}}
    \subfigure[\ $g^2(\omega,\kappa,p)$, $\kappa=1.5$]
    {\label{f:Fig4b}
    \includegraphics[width=0.45\linewidth]{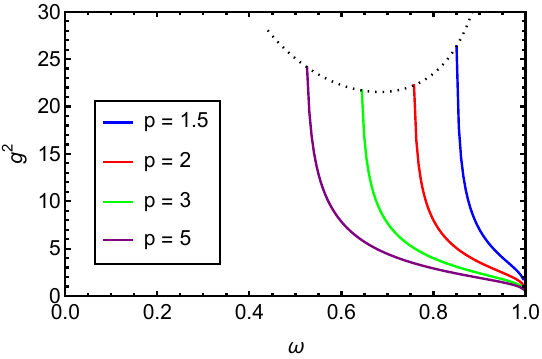}}
    \caption{\label{f:Fig4} Plots of $g^2(\omega,p)$ 
    when the charge $Q=1$ vs $\omega$ when (a) $p=2$ and
    (b) when $\kappa = 1.5$.  The dotted lines show the soliton domain.}
\end{figure}

For arbitrary $\kappa$, the charge is given by \eqref{e:QI-defs} so that if we 
set $Q=1$, the coupling constant $g$ is related to $\kappa$ and $p$ by the 
relation,
\begin{equation}\label{e:CaseII-gfun}
   g^{2}(\omega,\kappa,p)
   =
   \frac{2(\kappa+1) \, p \, \alpha}{\kappa^{\kappa} (\beta/m)^{\kappa-1} }
   \, I^{\kappa}(\omega,\kappa,p) \>,
\end{equation}
where $I^{\kappa}(\omega,\kappa,p)$ is given in \eqref{e:I-def}. Results are shown in Fig.~\ref{f:Fig4a} for $g^2(\omega,\kappa,p)$ vs $\omega$ for $p=2$, and in Fig.~\ref{f:Fig4b} for $g^2(\omega,\kappa,p)$ vs $p=2$ for $\kappa = 1.5$.  

\begin{figure}[t]
    \centering
    \subfigure[\ $p=2$]
    {\label{f:Fig5a}
    \includegraphics[width=0.45\linewidth]{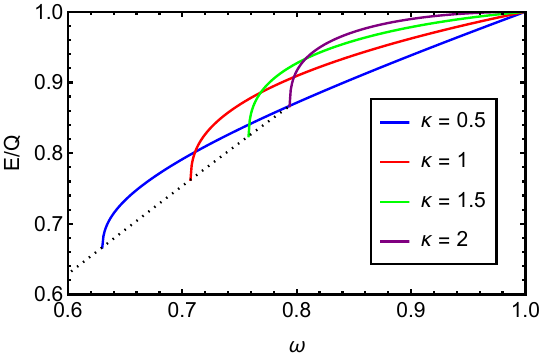}}
    \subfigure[\ $\kappa=1.5$]
    {\label{f:Fig5b}
    \includegraphics[width=0.45\linewidth]{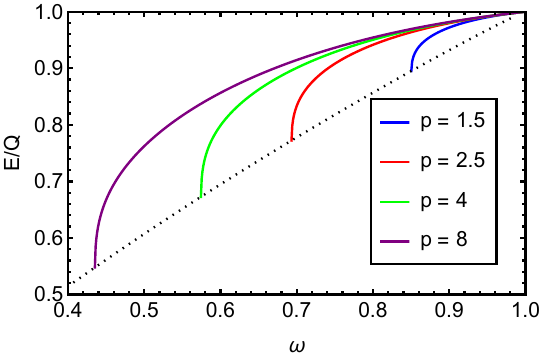}}
    \caption{\label{f:Fig5} Plots of $E/Q$ when (a) $p=2$ and
    (b) when $\kappa = 1.5$. The dotted lines show the soliton domain.}
\end{figure}

\begin{figure}[t]
    \centering
    \subfigure[\ $\kappa=0.9$]
    {\label{f:Fig6a}
    \includegraphics[width=0.45\linewidth]{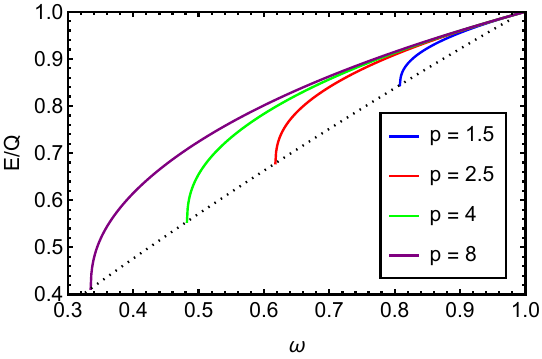}}
    \subfigure[\ $\kappa=3$]
    {\label{f:Fig6b}
    \includegraphics[width=0.45\linewidth]{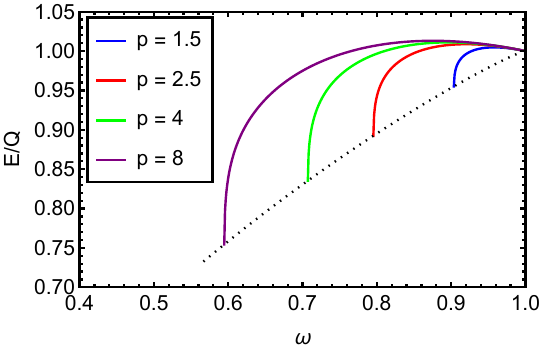}}
    \caption{\label{f:Fig6} Plots of $E/Q$ 
   vs $\omega$ when (a) $\kappa=0.9$ and
    (b) when $\kappa = 3 $. The dotted lines show the soliton domain.}
\end{figure}

We have the energy $E(\omega,\kappa,p)$ given by Eq.\eqref{Ekp}, namely 
\be 
E(\omega,\kappa,p) = \kappa  H_2 -(\kappa-1) \omega Q\,, 
\ee
so that 
\begin{equation}\label{e:E-Q1}
   E(\omega,\kappa,p)/Q(\omega,\kappa,p)
   =
   \kappa H_2(\omega,\kappa,p)/Q(\omega,\kappa,p) + (1-\kappa) \, \omega \>,
\end{equation}
where $H_2(\omega,\kappa,p)$ is given in \eqref{e:WJ-defs}. The ratio 
$E(\omega,\kappa,p)/Q(\omega,\kappa,p)$ is independent of $g$.
In Fig.~\ref{f:Fig5a} we plot $E/Q$ vs $\omega$ for $p=2$  and several values 
of $\kappa$, and in Fig.~\ref{f:Fig5b} we plot $E/Q$ vs $\omega$ for 
$\kappa=3/2$ and several values of $p$. As long as $\kappa<2$, we find that 
$E/Q$ is a monotonically increasing function of $\omega$  and $E/Q < 1$.
 In Fig.~\ref{f:Fig6a}, we plot $E/Q$ vs $\omega$ for $\kappa=0.9$ and in 
 Fig.~\ref{f:Fig6b}, $E/Q$ vs $\omega$ for $\kappa=3$.  In the last plot, 
 $E/Q > 1$ in a certain range of $\omega$, indicating possible instability. We 
 find that when $\kappa > 2 (E/Q)$  has a maximum somewhere between 
 $0.8 < \omega < 1$. At the maximum, $E/Q>1$.  This maximum persists all the way 
 to $p \rightarrow \infty$, which is the scalar-scalar theory.  We will discuss this further below.
%
%
%
%

%
%
%
%

%
%
%
%

\begin{figure}[t]
   \centering
   \includegraphics[width=0.75\linewidth]{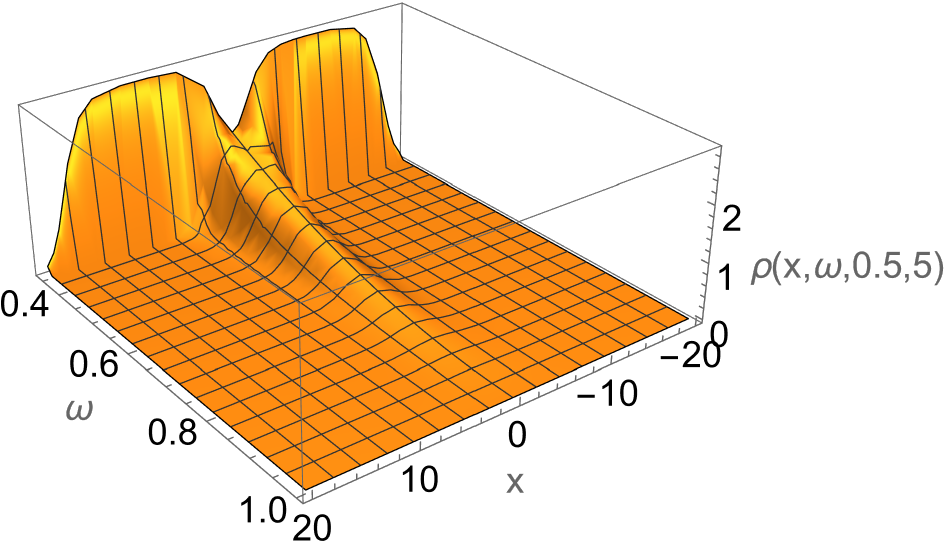}
   \caption{\label{f:Fig7}Plot of the charge density 
   $\rho(x,\omega,0.5,5)$ for the case when $\kappa=0.5$,  
   $p=5$, and $m=g=1$.  For this case, $\omega_{\text{min}} = 0.3420$}
\end{figure}

A plot of the density $\rho(x,\omega,\kappa,p)$ as a function of $x$ and $\omega$ where $\kappa = 0.5$ and $p = 5$ is shown in Fig.~\ref{f:Fig7}.  Plots with respect to other variables are quite similar.  
%
%
%
%

%
%
\subsection{\label{ss:VKstability}Vakhitov-Kolokolov stability}

\begin{figure}[t]
    \centering
    \subfigure[\ $Q$, $p=5$]
    {\label{f:FigVKQ}
    \includegraphics[width=0.45\linewidth]{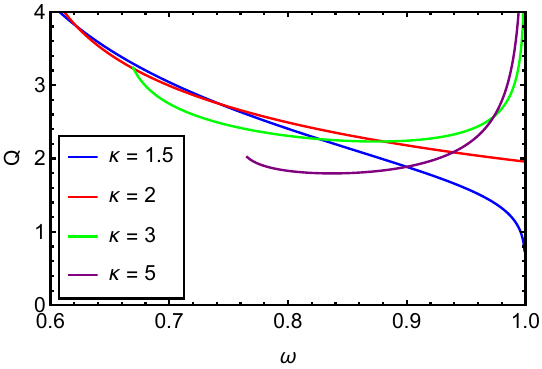}}
    \subfigure[\ $\dd{Q}/\dd{\omega}$, $p=5$]
    {\label{f:FigVKdQdomega}
    \includegraphics[width=0.45\linewidth]{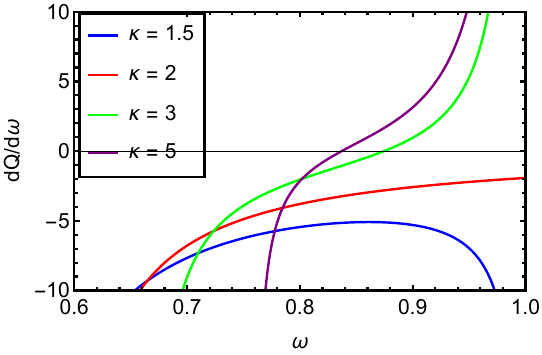}}
    \caption{\label{f:FigVKQQ} Plots of $Q$ and $\dd{Q}/\dd{\omega}$
    when $p=5$. Here $\dd{Q}/\dd{\omega} < 0$ for $\kappa <2$.} 
\end{figure}

The Vakhitov-Kolokolov stability criterion \cite{vk} is a condition for the 
linear stability (sometimes called the spectral stability) of the solitary wave
solutions to a wide class of $U(1)$-invariant Hamiltonian systems. The 
condition for the linear stability of a solitary wave with frequency $\omega$ 
has the form
\begin{equation}\label{e:dQdomega}
   \pdv{Q(\omega,\kappa,p)}{\omega} < 0 \>,
\end{equation}
where $Q(\omega)$ is the charge which is conserved by Noether's theorem due to 
the $U(1)$-invariance of the system. For the \smod\ system it is known that for
$\kappa >2$  there is a region $0 < \omega < \omega^\ast$ where 
$\partial Q(\omega)/\partial \omega < 0$. So this should also be true at least 
for large values of $p$. For our case, from \eqref{e:QI-defs} we have
\begin{equation}\label{e:VK-Q-I}
   \pdv{Q(\omega,\kappa,p)}{\omega}
   =
   \pdv{C(\omega,\kappa,p)}{\omega} \, I(\omega,\kappa,p)
   +
   C(\omega,\kappa,p) \, \pdv{I(\omega,\kappa,p)}{\omega} \>,    
\end{equation}
where
\begin{subequations}\label{e:dCdI}
\begin{align}
   \pdv{C(\omega,\kappa,p)}{\omega}
   &=
   - 
   \frac{2 [\,p(\kappa+1) \alpha \,]^{1/\kappa} \,[\,m - (\kappa-1) \, \omega \,]}
        { g^{2/\kappa} \, \kappa^2 \, \beta^{3-1/\kappa} } \>,
   \label{e:dCdomega-value} \\
   \pdv{I(\omega,\kappa,p)}{\omega}
   &=
   -
   \frac{2 \,  \, m}{\kappa \,  (m + \omega)^2}
   \int_{0}^{1} \!\!\dd{y}
   \frac{y^2}{(1 -y^2)^{(\kappa-1)/\kappa}} 
   \notag \\
   & \hspace{3em}
   \times \frac{(1 + \alpha^2 y^2)^{\kappa+1} 
      + p\, (1 - \alpha^2 y^2)^{\kappa})\,(1 + \alpha^2 y^2 + 2 \, \kappa) }
        { [\, p\, (1 - \alpha^2 y^2)^{\kappa+1} -
                  (1 + \alpha^2 y^2)^{\kappa+1} \,]^{(\kappa+1)/\kappa} } \>.
   \label{e:dIdpmega-value}
\end{align}
\end{subequations}
Results of the numeric integration are shown in Fig.~\ref{f:FigVKQQ} for the 
case when $m=1$, $g=1$, and $p=5$. Note that $\dd{Q}/\dd{\omega} < 0$ for 
$\kappa <2$. For $\kappa > 2$, there is a small region for $\omega_{\text{min}}
< \omega < \omega_c$ where $\dd{Q}/\dd{\omega} < 0$.  From 
Fig.~\ref{f:FigVKdQdomega}, the critical value $\omega_c$ where 
$\dd{Q}/\dd{\omega}= 0$ is seen to \emph{decrease} for increasing $\kappa$.

%
%

\subsection{Another Possible Indicator of Instability}
\begin{figure}
	\includegraphics[width=0.7\linewidth]{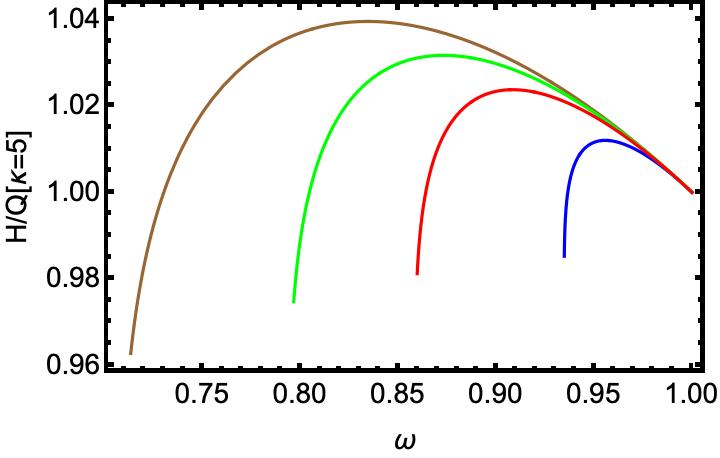}
	\caption{$H(\omega)/Q(\omega)$  for $\kappa=5$ and $p =1.5 $ (blue), $p =4$ (red), $p =2$ (green) and $p=8$ (brown).} 
	\label{EQk5}
\end{figure}

\begin{figure}
	\includegraphics[width=0.7\linewidth]{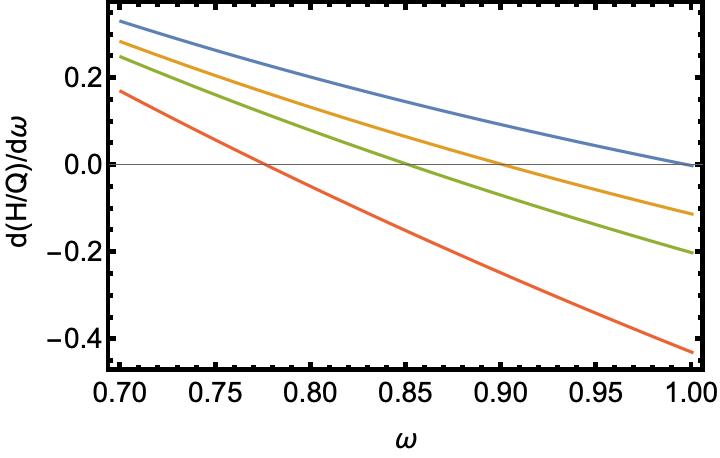} 
	\caption{$ \frac{d (H/Q)}{d \omega} $  for $p\rightarrow \infty$ and  $\kappa =2$ (blue), $\kappa =2.5 $ (yellow), $\kappa =3$  (green) and $\kappa=5$(red). } 
	\label{dhqdomss} 
\end{figure}

If we study the behavior of $E/Q$ as a function of $ \omega$ for various $p$, 
we find that once $\kappa > 2$, $E/Q$ has a maximum at finite omega and at the 
maximum $E/Q > 1 (m = 1)$.  To the right of the maximum $E/Q $ is a decreasing 
function of $\omega$. We believe that in this region the solutions are 
unstable.  This will have to be verified by numerics which will be a subject of
a forthcoming paper. An example of this behavior when $\kappa=5$ is shown 
in Fig.~\ref{EQk5}. We notice that the allowed region of $\omega$ for a 
solution to exist increases as we increase $p$. When $p \rightarrow \infty$, 
this model becomes the S-S model we studied earlier. For that model both $E$ 
and $Q$ can be obtained analytically. Further, one can also determine 
$\frac{d(E/Q)}{d \omega}$ analytically and plot where the maximum of the 
function occurs (i.e., when the derivative is zero). For this case we now plot 
the derivative at various $\kappa \ge 2$ so we can observe this behavior. We see 
from Fig. \ref{dhqdomss} that for the S-S case, a maximum first occurs when
$\kappa=2$. As we increase $\kappa$ the maximum occurs at decreasing $\omega$ 
giving a larger region in $\omega$ where the  derivative is negative. 
So the picture one gets at finite  $p$ is similar to Fig.~\ref{dhqdomss} but
the range of $\omega$  where this curve exists gets narrower as one decreases 
$p$ for fixed $\kappa$ since the solitary wave solutions only exist if 
$\omega > 1/p^{1/(\kappa+1)}$.
 
%

%
\section{\label{s:NonrelReduction}Nonrelativistic Reduction}

\begin{figure}[t]
   \centering
   \includegraphics[width=0.75\linewidth]{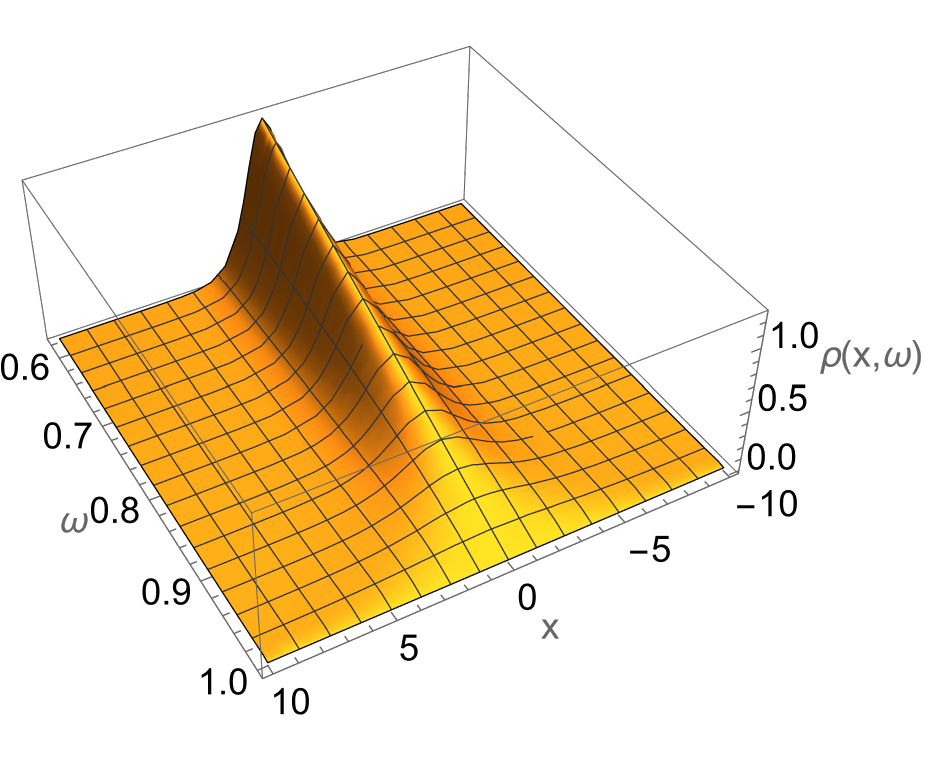}
   \caption{\label{f:Fig8}Plot of the charge density 
   $\rho_{\text{NR}}(x,\omega)$ for the non-relativistic reduction 
   for $\kappa=1$, $g=1$, and $p=3$.  Compare with the relativistic case
   in Fig.~\ref{f:Fig3}.}
\end{figure}

In \cite{fred} we showed that both the \vmod\ and \smod\ interactions to leading
order in $(m-\omega)/(2m)$ lead to the same modified nonlinear Schr\"odinger
\ equation.  Namely, letting 
\begin{equation}\label{e:NonRelPsi}
   \Psi(x,t) = \mqty(u(x)\\v(x)) \, \rme^{-\rmi \omega t} \>,
\end{equation}
we find that in leading order, $u(x)$ for both the \smod\ and \vmod\ obey the 
NLSE given by:
\begin{equation}\label{e:NonRelSch}
   \qty[\,
      - \partial_x^2 + m^2 - \omega^2 
      - g^2 \, (m + \omega) \, |u|^{2\kappa} \, ] \, u = 0 \>,
\end{equation}
which has the exact solution,
\begin{eqnarray}\label{e:NonRelExactSol}
   u(x) = A \sech^{1/\kappa}(\,\kappa \beta x \,)
   \qc
   A^{2\kappa} = \frac{(\kappa+1) \,(m-\omega)} {g^2}  \,. 
   \qc
\end{eqnarray}
The non-relativistic version of the generalized \ABS\ model then leads to:
\begin{equation}\label{e:ABSNonRelSch-I}
   \qty[\,
      - \partial_x^2 + m^2 - \omega^2 
      - \frac{p-1}{p} \, g^2 \, (m + \omega) \, |u|^{2\kappa} \, ] \, u = 0 \>,
\end{equation}
so the only effect of the additional $\vmod$ coupling is the change in the 
effective coupling constant to: $g^2 \rightarrow g^2 \, (p-1)/p$, in which case
\begin{equation}\label{e:ABSNonRelSch-II}
   A^{2\kappa}
   \rightarrow
   \frac{p}{(p-1)} \, \frac{(\kappa+1) \,(m-\omega) }{g^2 } \>.
\end{equation}
The non-relativistic density then becomes:
\begin{equation}\label{e:ABSNonRelrho}
   \rho_{\text{NR}}(x)
   =
   \qty[ \frac{p}{(p-1)} \, \frac{(\kappa+1) \,(m-\omega)}{g^2 } ]^{1/\kappa}
   \sech^{2/\kappa}(\,\kappa \beta x \,) \>.
\end{equation}
A plot of $\rho_{\text{NR}}(x)$ for the case when $\kappa=1$, $g=1$, and $p=3$ 
is shown in Fig.~\ref{f:Fig8}, which is to be compared with the relativistic 
result in \eqref{e:Requ-II}.  Although numeric values are close, the 
non-relativistic density lacks the double hump feature of the relativistic case.

%
%
%
%

%
%
\subsection{\label{ss:caseStability}Stability in the non-relativistic regime}

As shown above, for the generalized ABS case, to the leading order in 
$(m-\omega)/(2m)$ the only effect of the parameter $p$ is to 
change the coupling constant. However, if we go to the next order then the 
Hamiltonian of the mNLSE gets changed. For example, as shown in \cite{fred}, 
for the S-S case the modified Hamiltonian is given by
\be\label{5.7}
 H_{SS} = \int \frac{dx}{2m}[\psi^{*}_x \psi_x (1 
 + \frac{\hat{g}_{s}^2}{2m}(\psi^{*} \psi)^\kappa)] 
- \frac{\hat{g}_s^2}{\kappa+1} (\psi^{*} \psi)^{\kappa+1}\,, 
\ee
whereas for the V-V case the modified Hamiltonian is given by
\be\label{5.8}
 H_{VV} = \int \frac{dx}{2m}[\psi^{*}_x \psi_x (1 
 - \frac{\hat{g}_{v}^2}{2m}(\psi^{*} \psi)^\kappa)] 
- \frac{\hat{g}_{v}^2}{\kappa+1} (\psi^{*} \psi)^{\kappa+1}\,. 
\ee
Note that the next order correction has different signs for
the S-S and the V-V cases. On following the same procedure, one finds that in the
generalized ABS case, in the next order one does not just get a redefinition of
the coupling constant.  Instead one has:
\begin{equation}\label{e:NR-HVVmod}
   H_{\text{\ABS}}
   = 
   \int \frac{\dd{x}}{2m} 
   \qty{
   \qty[
      \psi^\star_x \psi_x 
      \qty(
         1 + \frac{\hat{g}^2 (1+1/p)}{2m} (\psi^\star \psi)^\kappa
      ) 
   ] 
   - 
   \frac{\hat{g}^2(1-1/p)}{\kappa+1} (\psi^\star \psi)^{\kappa+1}   
   } \>.
\end{equation}
Since $p>1$, the non-relativistic Hamiltonian of the generalized ABS model is 
the sum of two positive and one negative term:
 \be
 H_{\text{\ABS}} = H_1 + H_2 - H_3. \ee
   The action for this model is then
\begin{equation}\label{e:NRAction}
   S
   =
   \tint \dd{t} L[\psi,\psi^{\ast}]
   \qc
   L = \frac{\rmi}{2} \tint \dd{x} 
   (\, \psi^{\ast} \psi_t - \psi_t^{\ast} \psi \, ) - H_{\text{\ABS}} \>.
\end{equation}
In the non-relativistic regime where the mNLSE is valid we can use Derrick's 
theorem \cite{derrick} to study the stability with respect to scale 
transformations keeping the mass
\be
 M= \int_{-\infty}^\infty~ dx \abs{\psi} ^2
 \ee
 of the soliton fixed.  It is well known that for the NLSE, this method is a 
 reliable tool in determining the regions of instability. Derrick's theorem 
 states that if we make the transformation
\begin{equation}\label{e:NRDerrick}
   \psi(x) \rightarrow \beta^{1/2} \psi(\beta x) \>,
\end{equation}
which preserves the normalization, then if the energy is at a minimum, the 
system is stable. The effect on the energy is then given by:
\begin{equation}\label{e:NRDerrick-energy}
   H(\beta) = \beta^2 H_1 + \beta^{2+\kappa} HG_2 - \beta^{\kappa} H_3 \>.
\end{equation}
Setting the first derivative to zero at $\beta=1$ yields:
\begin{equation}\label{e:NRDerrick-firstderiv}
   2 H_1 + (2+\kappa) H_3 - \kappa H_3 = 0 \>,
\end{equation}
which is consistent with the equation of motion. To determine when $\beta=1$ is
a local minimum, we consider the second derivative at $\beta=1$, which is given by
\begin{align}
   \pdv[2]{H}{\beta} \Big |_{\beta=1}
   &=
   2 \, H_1 + (\kappa+2)(\kappa+1)\, H_2 - \kappa(\kappa-1) \, H_3
   \notag \\
   &\approx
   2 (2 - \kappa)\, H_1 + 2 (2 + \kappa)\, H_2 \>.
   \label{e:NRDerrick-secondderiv}
\end{align}
Since $H_2$ is positive and small relative to $H_1$, we find that up to and 
including $\kappa=2$, when the non-relativistic approximation is valid (i.e. 
when $(m-\omega)/(2m) \ll 1$), the system should be stable. 

%
\subsection{\label{ss:blowup}Blowup and critical Mass}

For the NLSE the Vakhitov-Kolokolov condition for stability is:
\begin{equation}\label{e:NR-VKstability}
   \pdv{M(\omega)}{\omega} > 0
   \qc
    M(\omega) = \tint \dd{x} |\psi_{\omega}(x)|^2  \,. 
\end{equation}
For our generalized NLSE we have for our approximate wave function
\begin{equation}\label{e:M-approximate}
    M(\omega)
    =
    \frac{\sqrt{\pi} \, \Gamma(1/\kappa)}
         { \kappa \, \Gamma(1/2+1/\kappa) \, (m^2 - \omega^2)^{1/2} }
    \qty[\, \frac{(\kappa+1) \, p \, (m -\omega)}
                { g^2 \, (p-1)} \,]^{1/\kappa} \>,
\end{equation}
so that
\begin{equation}\label{e:dMdomega-II}
   \pdv{M(\omega)}{\omega}
   =
   -
   \frac{\sqrt{\pi} \, \Gamma(1/\kappa) \, 
             [\, m + (1 - \kappa) \, \omega \,]}
        { \kappa^2 \, \Gamma(1/2+1/\kappa) \, (m^2 - \omega^2)^{3/2} }
    \qty[\, \frac{(\kappa+1) \, p \, (m -\omega)}
                { g^2 \, (p-1)} \,]^{1/\kappa} \>.
\end{equation}
Thus, the Vakhitov-Kolokolov stability will be satisfied if
\begin{equation}\label{e:VK-stability-condition}
   \omega \, ( \kappa - 1) < m = \omega \, (1 + \delta)
   \qq{$\Rightarrow$}
   \kappa < 2 + \delta \>.
\end{equation}
Here $\delta = (m - \omega)/\omega$, which is assumed small. So again we have 
that $\kappa < 2$ should be stable.  

We can follow Sec. V  of  Ref. \cite{fred} and use a self-similar variational 
approach to find that when $\kappa=2$ the solutions can blowup once the initial
mass of the soliton for the mNLSE is greater than a critical value. This 
follows from using the mNLSE Lagrangian. 

%
%
\section{\label{s:Conclusions}Conclusions}

In this paper we have generalized the ABS model \cite{abs} to a two (continuous)  
parameter  family of models characterized by the nonlinearity parameter 
$\kappa > 0$ and admixture of V-V vis a vis S-S parameter $p > 1$ and have 
obtained their solitary wave solutions. Unlike the S-S and the V-V cases, one of 
the novel features of these solutions is that the frequency $\omega$ is 
restricted to $1/p^{1/(\kappa+1)} < \omega/m < 1$. We have also shown that the
solitary wave bound states exist in the entire $(\kappa$-$p)$ plane ($\kappa > 0,
p>1$). Using the Vakhitov-Kolokolov criterion \cite{vk} we have further shown that these 
solutions are not only stable for $\kappa < 2$ but even for a small window 
beyond $\kappa > 2$. Finally, we have also considered the non-relativistic 
reduction of the generalized ABS model and shown that the corresponding 
generalized mNLSE solutions are stable in case $\kappa < 2$. This paper raises 
several questions, some of which are

\begin{enumerate}

\item  One would like to know the parameter range in
the $(\kappa$-$p)$ plane for which the solitary wave bound state solutions are 
stable. We would like to know if the stability is linked with a single/double 
hump solution and whether the stability is linked to if $E/Q > 1$. 

\item ABS \cite{abs} have also considered the PT-invariant generalization of 
their model. The interesting question is whether one can also construct 
the PT-invariant variant of our generalized ABS model and, if yes, can one also
obtain its exact solutions.

\item There has been a lot of discussion in the literature about the behavior 
of the S-S as well as the V-V solitary waves in various external fields 
\cite{mertens, fred1}. It is then natural to examine the behavior of the 
solitary waves of the generalized ABS model under various external fields.

\item In this paper we have considered two parameter families of models with
novel admixture of the V-V and the S-S interactions. Another possible interaction
is the PS-PS (pseudoscalar-pseudoscalar) interaction. Can one construct NLD 
models with the admixture of the PS-PS with the V-V and/or with the S-S models? 
Going further, can one construct a NLD model with an admixture of all three 
(i.e., S-S, V-V and PS-PS) interactions?

\item Finally, perhaps the most interesting question is if one can find 
relevance of the generalized ABS model (at least for few of the allowed $p$ 
values) in the context of some physical phenomena, for example in the 
Bose-Einstein condensate or some other physical system.
		
\end{enumerate}

We hope to address some of these questions in the near future.

\acknowledgments
We would like to thank Andrew Comech and Esthasios Charalampidis for useful 
discussions about stability. 
One of us (AK) is grateful to the Indian National Science Academy (INSA) for 
the award of the INSA Honorary Scientist position at Savitribai Phule Pune 
University, India. The work of AS was supported by the US Department of Energy. 

%
 
%
%
\end{document}